\documentclass[11pt,twocolumn,letterpaper]{article}

\usepackage{authblk}
\usepackage{graphicx}
\usepackage[margin=1in]{geometry}
\usepackage{multicol}
\usepackage{blindtext}

\let\OLDthebibliography\thebibliography
\renewcommand\thebibliography[1]{
  \OLDthebibliography{#1}
  \setlength{\parskip}{0pt}
  \setlength{\itemsep}{0pt plus 0.3ex}
}

\title{Mechanical Ringdown Studies of Large-Area Substrate-Transferred GaAs/AlGaAs Crystalline Coatings}

\author[1,*]{Steven D. Penn}
\author[2]{Maya M. Kinley-Hanlon}
\author[2, 3]{Ian A. O. MacMillan}
\author[4]{Paula Heu}
\author[4]{David Follman}
\author[5]{Christoph Deutsch}
\author[4, 5]{Garrett D. Cole}
\author[2]{Gregory M. Harry}

\affil[1]{Department of Physics, Hobart and William Smith Colleges, 300 Pulteney Street, Geneva, NY 14456, USA}
\affil[2]{Department of Physics, American University, 4400 Massachusetts Avenue, Washington, DC 20016, USA}
\affil[3]{Department of Physics, Georgetown University, 3700 O Street NW, Washington, DC 20057, USA}
\affil[4]{Crystalline Mirror Solutions LLC, 114 East Haley Street Suite G, Santa Barbara, CA 93101, USA}
\affil[5]{Crystalline Mirror Solutions GmbH, Lehargasse 1, A-1060 Vienna, Austria}

\affil[*]{Corresponding author: penn@hws.edu}

\begin{document}
\twocolumn[
  \begin{@twocolumnfalse}
\maketitle

\begin{abstract}
We investigated elastic loss in GaAs/AlGaAs multilayers to help determine the suitability of these coatings for future gravitational wave detectors. We measured large ($\approx 70$-mm diameter) substrate-transferred crystalline coating samples with an improved substrate polish and bonding method. The elastic loss, when decomposed into bulk and shear contributions, was shown to arise entirely from the bulk loss, $\phi_{\mathrm{Bulk}} = (5.33 \pm 0.03)\times 10^{-4}$, with $\phi_{\mathrm{Shear}} = (0.0 \pm 5.2) \times 10^{-7}$. These results predict the coating loss of an 8-mm diameter coating in a 35-mm long cavity with a 250-$\mu$m spot size (radius) to be $\phi_{\mathrm{coating}} = (4.78 \pm 0.05) \times 10^{-5}$, in agreement with the published result from direct thermal noise measurement of $\phi_{\mathrm{coating}} = (4 \pm 4) \times 10^{-5}$. Bonding defects were shown to have little impact on the overall elastic loss.
\end{abstract}
\end{@twocolumnfalse}]
\section{Introduction}
\label{sec:Intro}
The direct measurement of gravitational waves~\cite{abbott061101,abbott241103}, as predicted by Einstein's General Theory of Relativity, has opened a new window on the universe and launched the field of multi-messenger astronomy~\cite{abbott848L13,ashton2018, goldstein2017}. Interferometric gravitational wave detectors, such as LIGO~\cite{abbott2016}, Virgo~\cite{acernese2014}, and KAGRA~\cite{kagra} are precision optical instruments designed specifically to detect these distortions in space-time. Thermally-driven fluctuations of the optical coatings of the detector test mass mirrors are a significant limitation to their sensitivity~\cite{abbott2016} and astronomical reach. Reducing coating thermal noise to near or below the standard quantum limit~\cite{khalili,heurs,zhou} is a key goal for future detectors~\cite{abbott2018}. Thermal noise is expected to limit the sensitivity of the current LIGO and Virgo detectors~\cite{abbott2016, acernese2014} in the mid-frequency band, 50-150~Hz, which is their region of highest sensitivity~\cite{abbott2016}. In addition, thermal noise will present a significant challenge when designing future, more sensitive, gravitational wave detectors~\cite{abbott2017b}. Ultimately, the minimization of thermal noise will allow for fully quantum-limited interferometry.

crystalline coatings have shown promising results, with sub-ppm absorption and scatter in-line with that seen in ion-beam sputtered coatings~\cite{marchi} currently employed in gravitational wave detectors~\cite{abbott2016}. These promising optical properties now motivate us to explore the thermal noise performance of large-area crystalline coatings.

Epitaxial GaAs/Al$_{0.92}$Ga$_{0.08}$As (AlGaAs) multilayers have demonstrated low elastic losses in free-standing microresonator experiments at both room and cryogenic temperatures \cite{cole2008, cole2012}. Moreover, direct thermal noise measurements in reference-cavity-stabilized laser systems have confirmed the low loss of these single crystal films $\phi_{\mathrm{coating}} = (4 \pm 4) \times 10^{-5}$ once transferred to a final optical substrate and implemented as a high-reflectivity interference coating \cite{cole2013}. In addition to their low elastic loss, recent optical characterization efforts on large-area (50-mm diameter) crystalline coatings have shown promising results, with sub-ppm absorption and scatter in-line with that seen in ion-beam sputtered coatings \cite{marchi} as currently employed in gravitational wave detectors. In terms of size scaling, crystalline coatings may currently be manufactured with diameters up to 20 cm using commercially available wafers, with the possibility for realizing 40-cm diameter optics using custom-fabricated GaAs substrates. While these results are promising for future implementation in gravitational wave detectors, noise in these coatings have thus far been probed on either small-area optics (typical coating diameters from 5-8 mm) or with small spot sizes at the millimeter scale. In contrast, current LIGO test mass mirrors have 34-cm diameter faces \cite{billingsle} with cm-scale optical spots, and future gravitational wave interferometers may employ larger mirrors in part as a method to reduce coating thermal noise which depends inversely on the beam diameter.

Large beams require a uniform coating surface across the full face of the suspended optic. Any defects, even far from the beam’s center, could generate excess optical loss, thereby increasing shot noise and possibly increasing mechanical loss in the coating, reducing the detector sensitivity.  On the later point, in the course of developing larger AlGaAs mirror coatings, concerns have been raised that low elastic losses may be difficult to achieve at larger size scales. Variations in the dissipation mechanism due to imperfections or varying bond strength across a sample may allow for low losses to be achieved in small scale measurements, but not for increased sample sizes. For larger samples, the increased coating area also increases the likelihood of a bond defect occurring between the AlGaAs coating and the substrate. These defects could be expected to increase the elastic loss. 

We report here on elastic loss measurements on a set of two 70.1-mm diameter AlGaAs crystalline coatings (Samples 2 \& 4), with almost a factor of 80 larger coating area than previously investigated. In order to minimize potential interface losses the silica substrates were precision polished and efforts were made to optimize the GaAs-to-silica bond quality. After production, both samples exhibited about 10 visible defects and a few larger flaws along the edge. The elastic loss was measured for both samples using mechanical ringdown.

Following the initial loss measurement, Sample 4 was subjected to a selective chemical etching process to remove the bonding defects and was remeasured. The coating elastic loss was then calculated for the three sets of measurements (Samples 2, 4, \& 4 etched). We found that the coating loss before and after etching showed only minor differences, indicating that the bond defects did not contribute significantly to the loss. In addition, the coating elastic loss measured for Mode 1 was less than that measured in any previous experiments with AlGaAs, indicating no significant excess loss induced by interfacial defects. Finally the separate components of elastic loss extracted from these measurements predict a coating loss for a 35-mm optical reference cavity that is consistent with published values~\cite{cole2013}.

\section{Background}
\label{sec:background}
Brownian thermal noise can be described by the Fluctuation-Dissipation Theorem~\cite{callen1951, greene1951}, which demonstrates that the fluctuations in the state of a system and the system's dissipation can both be described by an elastic loss angle, the ratio of the imaginary part of the complex elastic constant to the real part. In 1998, Levin used the Fluctuation-Dissipation Theorem to calculate the contribution of thermal noise in the test mass mirror coatings to LIGO's overall sensitivity~\cite{levin1998}. This calculation was a revelation to the gravitational wave community, revealing that the thermal noise contribution from a few micrometers of lossy coating material could greatly exceed the thermal noise from the $> 10$ cm thick fused silica substrate. When Levin's derivation is applied to the case of amorphous mirror coatings on gravitational wave detector test masses, the coating thermal noise equation is given by~\cite{harry2002}:
\begin{equation}
\label{eqn:noise}
S_X(f) = 2 k_B T \phi_{\mathrm{eff}}\frac{1-\sigma^2}{\pi^{3/2} f w Y},
\end{equation}
where $S_x$ is the power spectral density of position fluctuations, $f$ is the frequency, $k_B$ is the Boltzmann constant, $T$ is the temperature, $\sigma$ is the Poisson ratio of the optic substrate material, $w$ is the half-width of the Gaussian mode of the laser, $Y$ is the Young's modulus of the optic substrate, and 
\begin{equation}
\label{eqn:phieff}
\phi_{\mathrm{eff}} = \phi + \phi_{\mathrm{coating}}\frac{2d-4 d \sigma}{\sqrt{\pi} w (1-\sigma)},
\end{equation}
where $\phi$ is the loss angle of the optic substrate, $d$ is the thickness of the coating, and $\phi_{\mathrm{coating}}$ is the loss angle of the coating. Equation~\ref{eqn:phieff} is a simplification of the full formula for $\phi_{\mathrm{eff}}$ assuming only a single loss angle $\phi_{\mathrm{coating}}$ and elastic constant $Y$ can be used to characterize the elasticity of the coating material. 

For our experiments, the samples consist of thin coating layers bonded to or deposited on thin substrates formed from a very low loss material, typically fused silica. The elastic loss of the sample may be determined by measuring the modal $Q$ factor via mechanical ringdown. This weakly-damped system can be driven to resonance and then allowed to freely ringdown with the amplitude describing a decaying exponential $A_0 e^{-t/\tau}$. The quality factor and elastic loss are related by $Q = \pi f_0 \tau = 1/\phi_{\mathrm{sample}}$ where $f_0$ is the resonant frequency of the normal mode.

The elastic loss angle is the fraction of energy dissipated during each oscillation. Therefore, one can extract the coating loss using
\begin{equation}
\label{eqn:phifromQ}
\phi_{\mathrm{coating}} = \frac{\phi_{\mathrm{sample}} - R_{\mathrm{substrate}} \phi_{\mathrm{substrate}}}{R_{\mathrm{coating}}},
\end{equation}
where $R_{\mathrm{substrate}} = E_{\mathrm{substrate}}/E_{\mathrm{sample}} \approx 1$ and $R_{\mathrm{coating}} = E_{\mathrm{coating}}/E_{\mathrm{sample}}$ are the energy ratios of the system components, also known as the dissipation dilution factors. The energy ratios, which were calculated using a finite element model, are provided in Table~\ref{table:Rvalues}. 

AlGaAs, which is a face-centered cubic crystal, has an equation of elasticity that is expressed, using Voight notation, as $\sigma_I = c_{IJ} \epsilon_J$, where the elasticity matrix, $c_{IJ}$ depends on three independent constants.
\begin{equation}
\label{eqn:c_IJ}
c_{IJ} = \left[     \begin{array}{cccccc}
        c_{11} & c_{12} & c_{12}  \\
       c_{12}& c_{11} & c_{12}  \\
       c_{12} & c_{12} & c_{11}  \\
       &&& c_{44}\\
       & & & & c_{44}\\
       & & & & & c_{44} \end{array}\right]
\end{equation}
In Eqn.~\ref{eqn:c_IJ} the unspecified elements are zero. For AlGaAs, $c_{11}=119.94$~GPa, $c_{12}=55.38$~GPa, and $c_{44}=59.15$~GPa. For each of the three elastic constants, there should be a unique loss angle, $\phi_{11}$, $\phi_{12}$, and $\phi_{44}$. The coating loss angle is then given by:
\begin{equation}
\phi_{\mathrm{coating}} = R_{\mathrm{11}} \phi_{\mathrm{11}} + R_{\mathrm{12}} \phi_{\mathrm{12}} + R_{\mathrm{44}} \phi_{\mathrm{44}},
\end{equation}
where $R_{xx} = E_{xx}/E_{\mathrm{coating}}$ and $E_{xx}$ is the elastic energy in the $xx$ deformation. ($11 =$ parallel stress-strain, $12 = $ orthogonal stress-strain, and $44=$ shear stress-strain). Because a single loss angle is measured for each mode of the sample, one determines the contributing loss angles by fitting the sample loss as a function of mode frequency. This method requires that the energy ratio functions be linearly independent in order to avoid degeneracy.

\begin{figure}[!htbp]
    \centering
    \includegraphics[width=.45\textwidth]{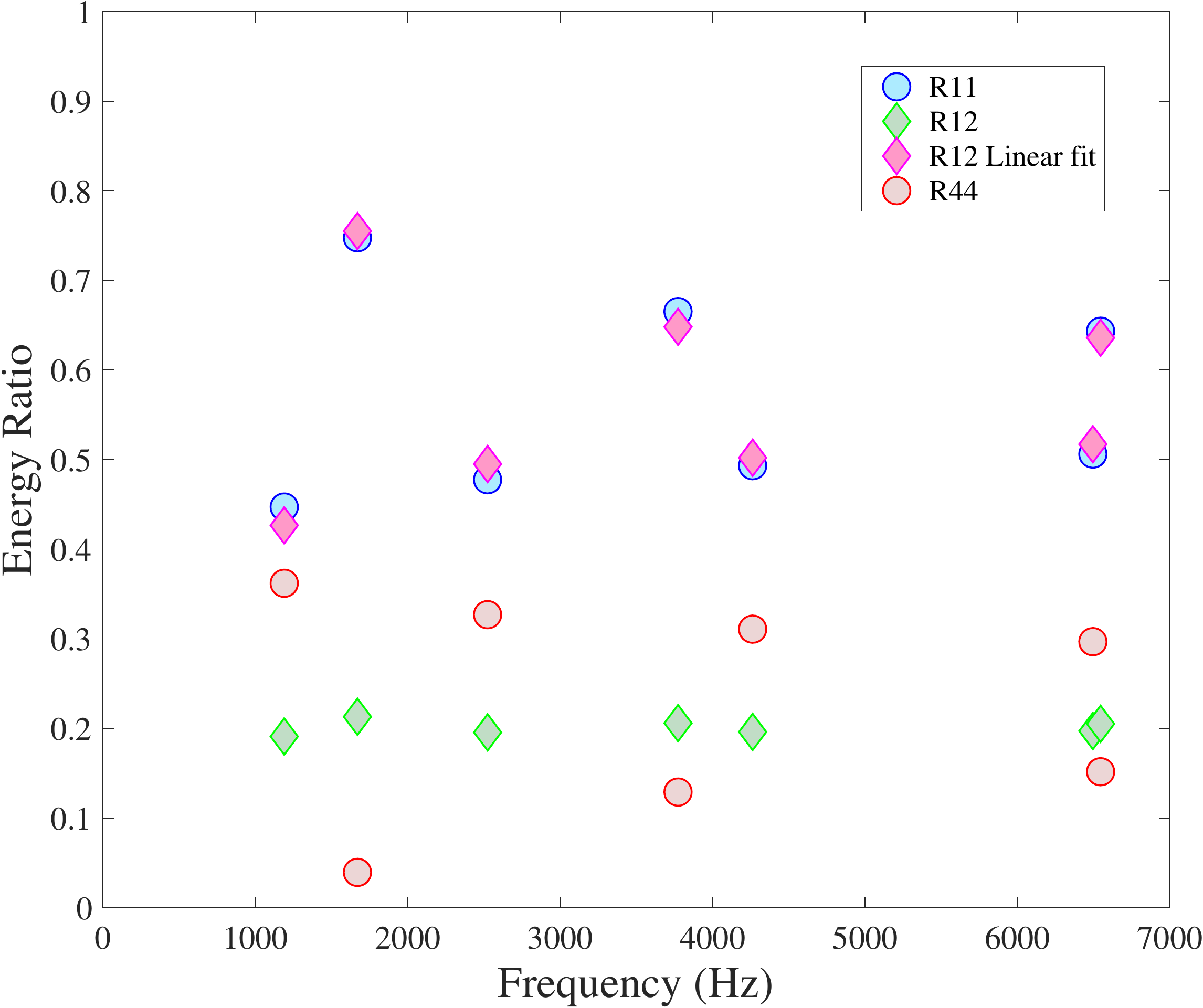}
    \caption{The energy ratios, $R_{11}$, $R_{12}$, and $R_{44}$, for the modes measured in Sample 4. The $R_{12}$ Linear Fit demonstrates the $R_{11}$, $R_{12}$ dependence since $R_{11} \approx m\cdot R_{12} + b$ }
    \label{fig:AnisoEnergy}
\end{figure}

Unfortunately $R_{11}$ and $R_{12}$ are similar functions of mode, which makes it difficult to distinguish $\phi_{\mathrm{11}}$ and $\phi_{\mathrm{12}}$.  
It should be possible to separate these loss angles by examining modes with quadrupole symmetry and comparing the loss of degenerate states that are aligned and unaligned with the crystal axes. However, for the modes we measured this was not possible and as can be seen in Figure~\ref{fig:AnisoEnergy}, $R_{11}$ and $R_{12}$ have a linear dependence.  

Therefore, for the remainder of this paper we will characterize the elastic loss in AlGaAs coatings using a bulk/shear decomposition, a method usually employed in analyzing the loss and thermal noise in amorphous coatings~\cite{hong,abernathy}. Bulk/Shear decomposition is an appropriate choice in this case because $R_{\mathrm{shear}} = R_{44}$ and $R_{\mathrm{bulk}}$ is composed from $ R_{11}$ and $R_{12}$. As will be shown in the Results Section~\ref{sec:results}, this choice appears to reflect a natural separation for AlGaAs coatings (see Figure~\ref{fig:bulkLoss}).

\begin{table*}[ht]
\centering
\begin{tabular}{ccccccc}
\hline
\centering
& \multicolumn{3}{c}{Initial Sample}& \multicolumn{3}{c}{Etched Sample}\tabularnewline
Mode &  $R_{\mathrm{coating}}$ &  $R_{\mathrm{bulk}}$&  $R_{\mathrm{shear}}$ & $R_{\mathrm{coating}}$ &  $R_{\mathrm{bulk}}$ & $R_{\mathrm{shear}}$ \tabularnewline
\hline 
1  &  0.0218  &  0.0347  &  0.965  &  0.0242  &  0.0381  &  0.962  \tabularnewline
2  &  0.0228  &  0.320  &  0.680  &  0.0216  &  0.315  &  0.685  \tabularnewline
3  &  0.0193  &  0.0568  &  0.943  &  0.0166  &  0.0584  &  0.942  \tabularnewline
4  &  0.0233  &  0.239  &  0.761  &  0.0219  &  0.251  &  0.749  \tabularnewline
5  &  0.0175  &  0.0724  &  0.928  &  0.0149  &  0.0698  &  0.930  \tabularnewline
6  &  0.0161  &  0.0839  &  0.916  &  0.0133  &  0.0827  &  0.917  \tabularnewline
7  &  0.0226  &  0.217  &  0.783  &  0.0189  &  0.243  &  0.758  \tabularnewline
\hline
\end{tabular}
\label{table:Rvalues}
\caption{Dissipation dilution factors, $R$, used to extract the coating loss and the bulk/shear components of the coating loss.}
\end{table*}

The high reflectivity coatings used currently in Advanced LIGO and Advanced Virgo are multilayers of amorphous metal oxides, with alternating layers of SiO$_2$ (low index) and TiO$_2$-alloyed Ta$_2$O$_5$ (high index) deposited by ion-beam sputtering~\cite{harry2007, harry2010, crooks, penn2003, reid}. These dielectric multilayer coatings exhibit excellent optical properties including $<$1 ppm of absorption and ppm-level scatter~\cite{billingsle} and can be applied over a large area on a variety of optical substrates. However, the main drawback of these coatings is the high elastic loss of the high index material, which generates unacceptably high levels of coating thermal noise. The Initial LIGO coatings were a SiO$_2$ / Ta$_2$O$_5$ quarter-wave multilayer coating with an elastic loss of $\approx 3 \times 10^{-4}$~\cite{penn2003}. For Advanced LIGO the coatings were improved by alloying the Ta$_2$O$_5$ layers with TiO$_2$, which reduced the coating loss to $\approx 2 \times 10^{-4}$~\cite{harry2007}.

Significant improvements in performance are still being investigated for low-noise and high-reflectivity coatings. Specifically, for the recently funded ``A+'' upgrade for Advanced LIGO, the goal is to reduce the coating elastic loss by another factor of 2--4~\cite{barsotti}.

Single-crystal interference coatings, such as GaAs/AlGaAs, are an attractive candidate for future gravitational wave detectors. This material simultaneously exhibits excellent optical quality~\cite{cole2016,marchi} and low elastic loss~\cite{cole2008,cole2012}, with a measured coating loss of $\phi_{\mathrm{coating}} = (4 \pm 4) \times 10^{-5}$ at room temperature when implemented in an optical reference cavity~\cite{cole2013}. These coatings consists of alternating lattice-matched single crystal films deposited via an epitaxial growth process. Al$_{x}$Ga$_{1-x}$As, 0$<x<$1, is a ternary alloy of GaAs and AlAs III-V compound semiconductors, both consisting of a face-centered cubic unit cell and a nearly matching lattice constant across all Al compositions. The ability to generate low-strain heterostructures with reasonable refractive index contrast allows for the generation of high-performance single-crystal optical interference coatings as initially demonstrated by van der Ziel and Illegems~\cite{vanderZiel}. One major engineering challenge to this material system, being a single-crystal structure, is the need for lattice matching in epitaxy, which precludes the growth of such heterostructures on arbitrary optical surfaces, including direct-deposition on amorphous or mismatched crystalline structures. To overcome this limitation, epitaxial multilayers are removed from their initial growth wafers and directly bonded to the final optical surface. With this approach, high purity and low defect density single crystal materials can be combined with arbitrary (including curved) optical substrates~\cite{cole2016}.

\section{Method}
\label{sec:method}
\begin{figure}[!htbp]
    \centering
    \includegraphics[width=.45\textwidth]{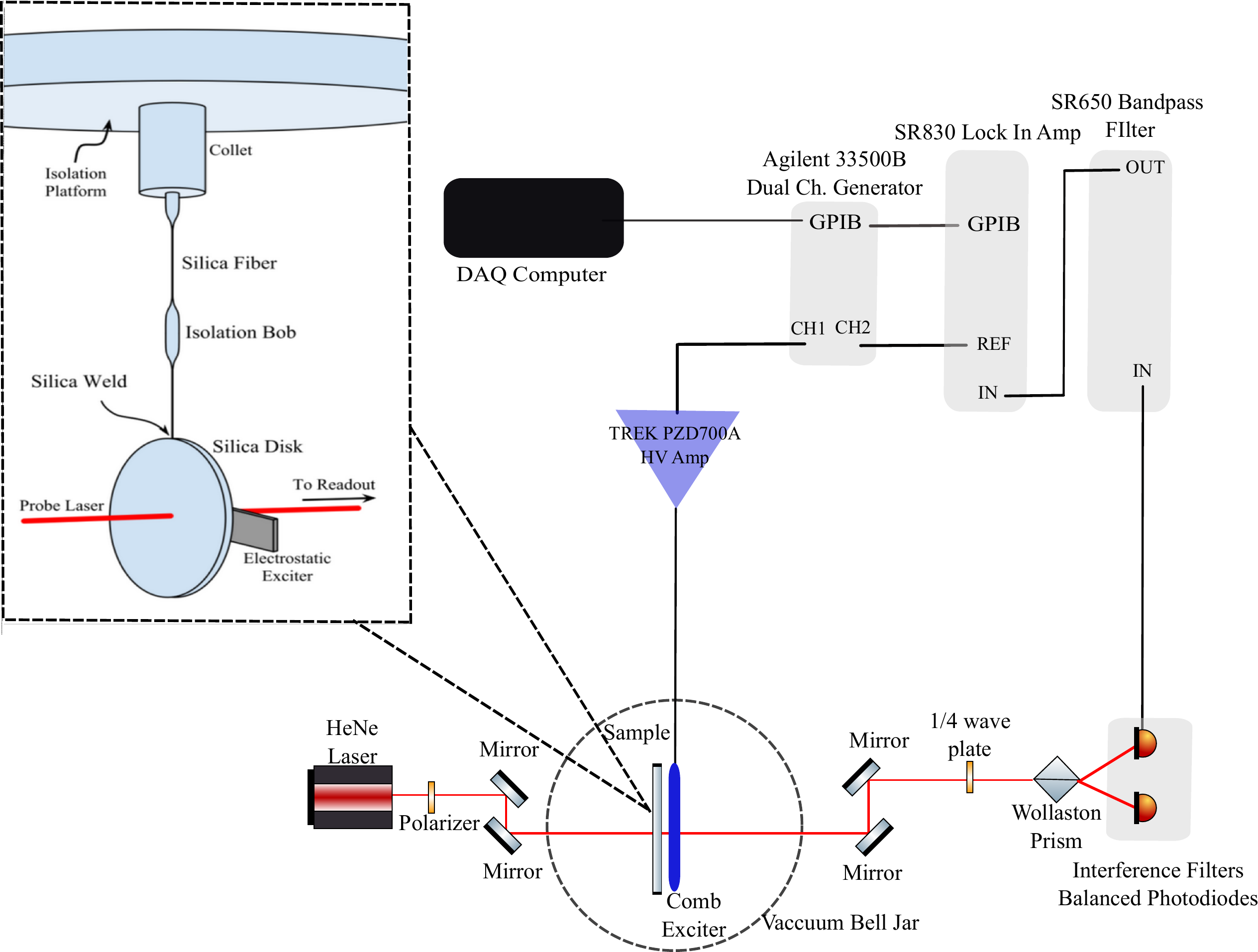}
    \caption{Experimental setup used to measure the elastic loss. The inset shows how the sample is hung with a thin silica fiber, connected with an isolation bob for vibration isolation. The comb capacitor, in blue, is situated close to the sample for efficient driving.}
    \label{fig:setup}
\end{figure}

The 3" diameter fused silica substrates (Corning 7980) employed in our experiment were obtained from a commercial wafer manufacturer with specifications of $<$0.5-nm RMS microroughness, a wafer bow/warp of $<$15 $\mu$m, and 1-mm thickness with a total thickness variation of $<$10 $\mu$m. Before measurement (and coating), Sample 4 was annealed at a maximum temperature of 950 °C for approximately 6 hours in a clean air atmosphere. Sample 2 was not annealed. Each sample was then suspended in a vacuum bell jar from a welded silica fiber suspension~\cite{harry2002}. A vacuum was maintained below $10^{-5}$~Torr throughout the measurement. This technique for measuring mechanical $Q$s has been described in several papers~\cite{kinleyhanlon,harry2002,harry2007,penn2003,crooks,reid}. We summarize the process in the text below and include a diagram of the experiment in Figure~\ref{fig:setup}. 

To excite the mechanical modes of the sample, a comb capacitor (exciter) was placed near the suspended sample (see the inset in Figure~\ref{fig:setup}). The exciter generates an alternating gradient electric field that exerts an oscillatory force on the induced dipole ($\vec{F} = \vec{p} \cdot \vec{\nabla}\vec{E}$). The sample is driven at the normal mode frequency. Excitation is ceased and the free decay is measured by recording the strain-induced birefringence (ellipsometry)~\cite{harry2002}. 
The data is heterodyned using a lock-in amplifier and recorded using a LabView data acquisition code written by the author (SP). A typical data run is recorded over a period of at least twice the exponential decay factor or the time it takes the amplitude to decrease by a factor of $e^{-2}$. Several data runs are recorded for each mode, and the loss for each mode is the average of the results weighted by the fit uncertainty assuming Gaussian statistics. The loss was measured for the bare substrate and for the coated samples. The coating loss was calculated using Equation~\ref{eqn:phifromQ}.

\begin{figure}[!htbp]
    \centering
    \includegraphics[width=.225\textwidth]{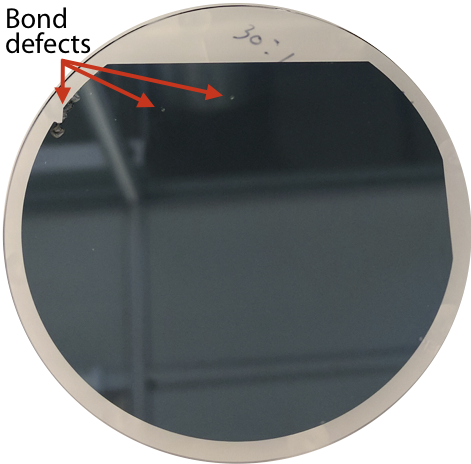}
    \includegraphics[width=.225\textwidth]{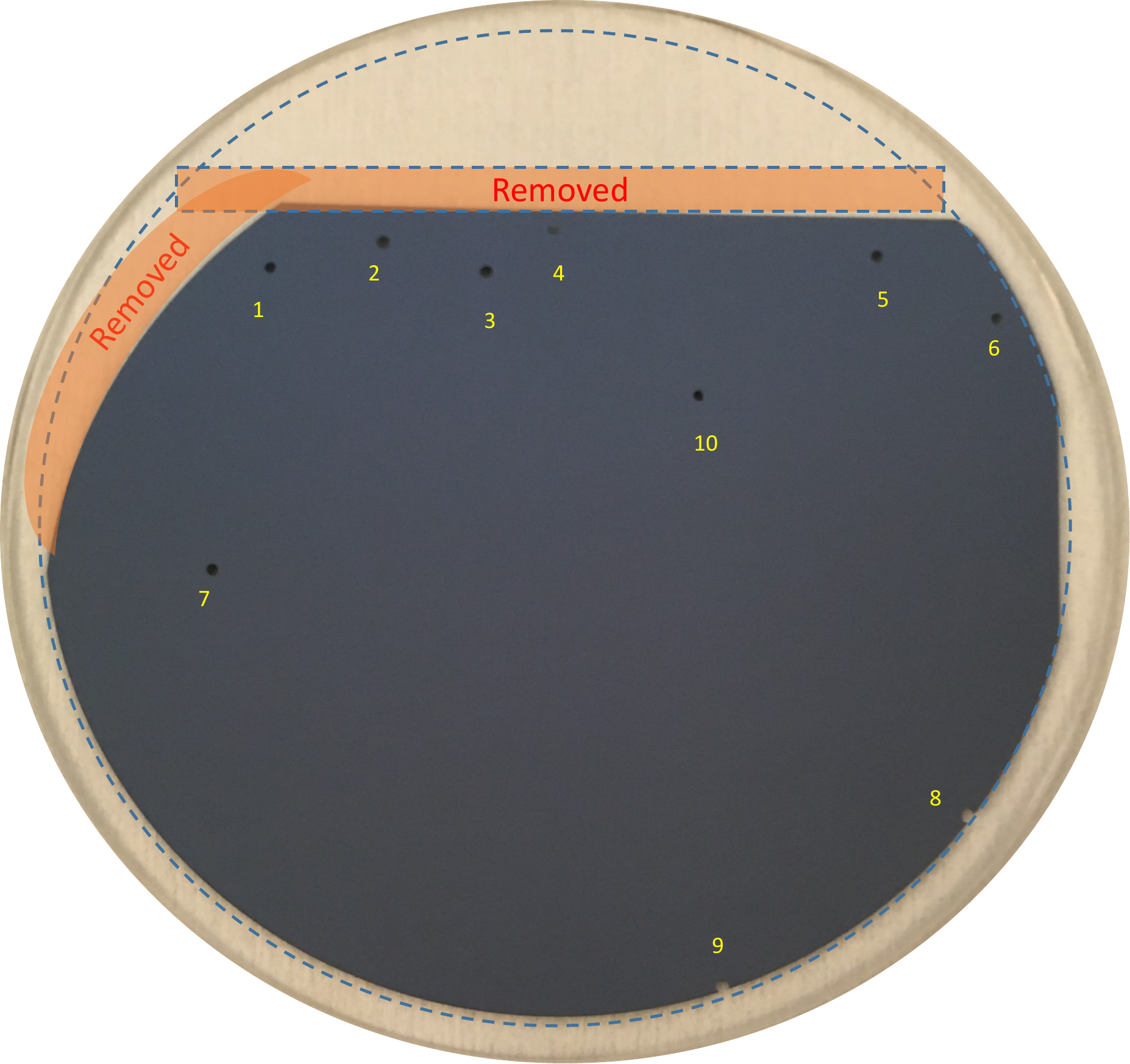}
    \caption{Photographs of the Sample 4 before (left) and after (right) the selective defect removal process. The etched regions have been highlighted in the right panel.}
    \label{fig:defects}
\end{figure}

After the measurement of the substrates, the samples were coated with a high-reflectivity AlGaAs multilayer with 35.5 layers of alternating GaAs (76.43~nm) and Al$_{0.92}$Ga$_{0.08}$As (89.35~nm), for a target optical transmission of 10 ppm at 1064 nm. Similar to previous crystalline coating efforts~\cite{cole2013,schreiber,chalermsongsak,cole2012,marchi}, we begin by growing a single-crystal multilayer by molecular beam epitaxy (MBE) on a 150-mm diameter GaAs wafer (in a 7" $\times$ 6" wafer configuration). For this effort, we deposit 36 layer pairs of quarter-wave (optical thickness at a wavelength of 1064 nm) GaAs/Al$_{0.92}$Ga$_{0.08}$As with the final Al-containing layer acting as an etch stop for selective substrate removal. Following the MBE growth process, each 6" wafer is lithographically patterned to generate two approximately 3" diameter coating discs with a large "flat" for crystal orientation identification as well as to pull back the coating from the heat-affected zone generated in fiber welding. These discs were inspected, thoroughly cleaned, and then directly bonded to a 3" diameter, 1-mm thick precision polished fused silica substrate. Following the substrate-transfer coating process, the mirror surface was again thoroughly cleaned and inspected for imperfections. Then the sample's $Q$ were remeasured at the same normal modes. Similar to the coating investigated in~\cite{marchi}, completed samples exhibited a small population of visible defects. The example shown in Figure~\ref{fig:defects} had imperfections at 12 locations, including point defects $>$50 $\mu$m in diameter, as well as larger, unbonded regions at the coating edge.

Following the measurements on the coated samples, Sample 4 was further processed to eliminate the macroscopic bond defects. First the mirror surface was covered with photoresist. Then, a filtered (short-wavelength blocking) white-light optical microscope was used to identify and expose, via removal of said filter, the applied photosensitive polymer film over each defect. After exposure, the mirror was submerged in a developer solution to remove the photoresist at the defect sites, and a selective phosphoric-acid based wet chemical etch (H$_{3}$PO$_{4}$:H$_{2}$O$_{2}$:H$_{2}$O 1:5:15) was used to remove the undesired coating material. Our experience has shown that this etch has high selectivity with SiO$_{2}$ and we can recover a pristine surface with sub-Angstrom RMS microroughness. When the etching process was complete, the loss in Sample 4 was measured again.

\begin{figure}[htbp]
    \centering
    \includegraphics[width=.45\textwidth]{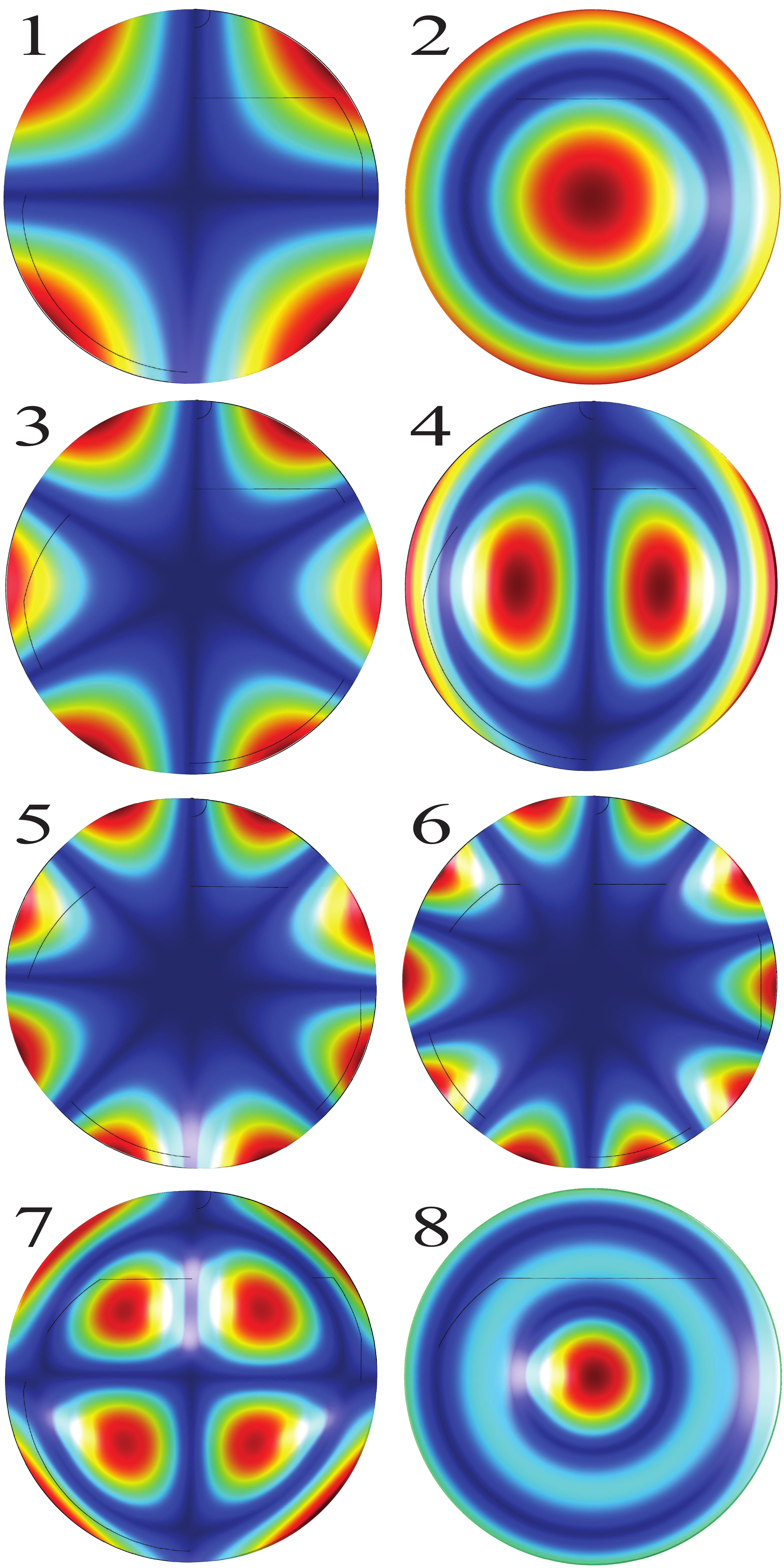}
    \caption{Modes 1--8 of the coating sample. The \textit{radial} modes (1, 3, 5, 6) are dominated by shear energy. The \textit{drumhead} modes (2, 4, 7, 8) have $1/3$ of their energy in bulk stress. Modes 5 and 8 are presented for completeness, but no data was collected on these modes.}
    \label{fig:Modes}
\end{figure}

\section{Results}
\label{sec:results}

\begin{figure} [t]
    \centering
    \includegraphics[width=.45\textwidth]{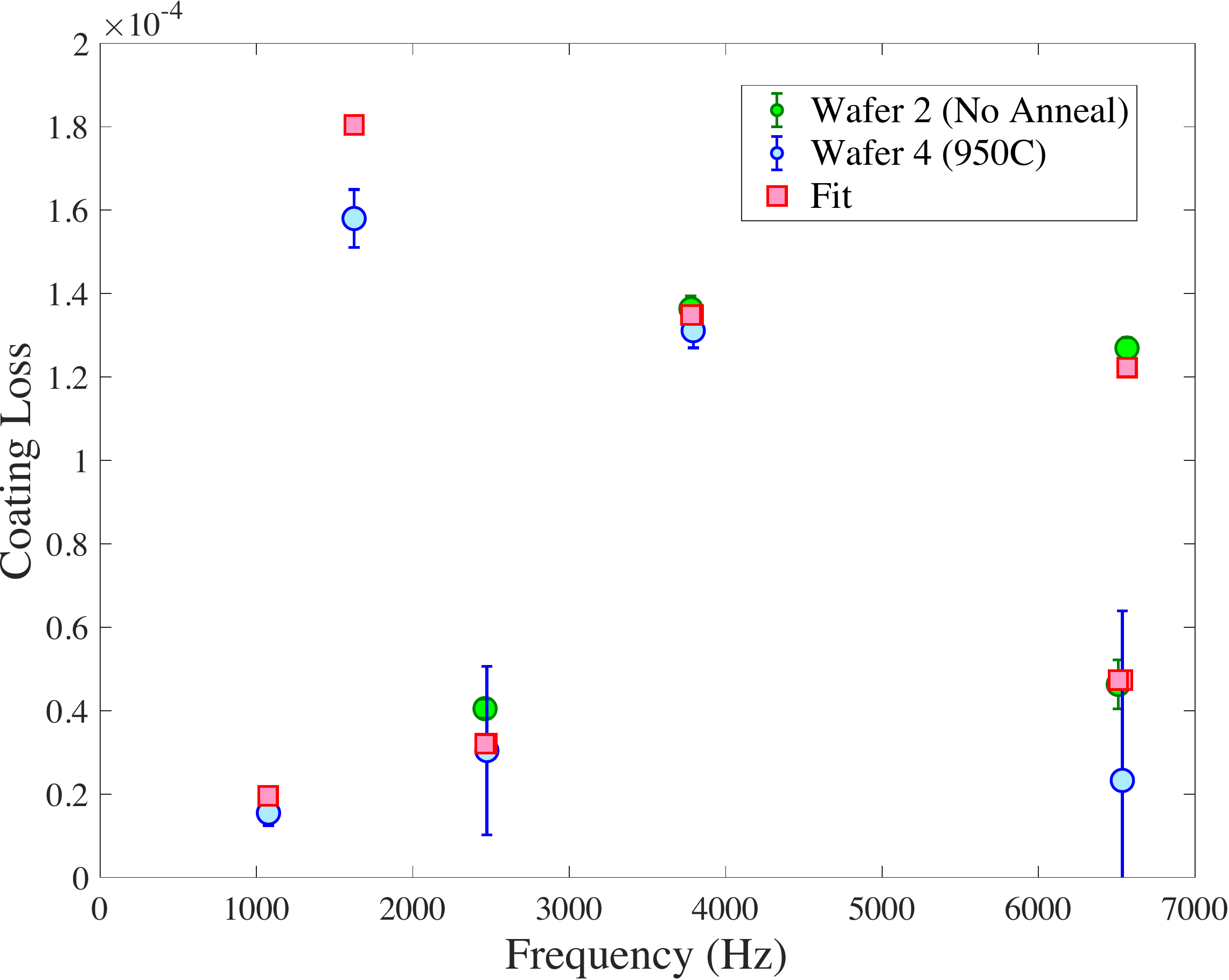}
    \caption{Coating loss of the initial samples with bond defects. A fit of bulk and shear losses yielded $\phi_{\mathrm{Bulk}} = (5.64 \pm 0.10)\times 10^{-4}$ and $\phi_{\mathrm{Shear}} = (0.0 \pm 1.5) \times 10^{-6}$. }
     \label{fig:initialLoss}
\end{figure}

\begin{table}[!t]
\begin{center}
\begin{tabular}{ccccc}\hline
Mode & Freq. & \multicolumn{3}{c}{Loss Angle $(\times 10^{-5})$}\tabularnewline
 &  (kHz) & $\phi_{\mathrm{sample}}$ & $\phi_{\mathrm{substrate}}$& $\phi_{\mathrm{coating}}$ \tabularnewline
\hline 
1  &  1.074  &  0.0955  &  0.0657  &   1.4400 \tabularnewline
3  &  2.462  &  0.1889  &  0.1131  &   4.0500 \tabularnewline
4  &  3.778  &  0.3351  &  0.0172  &  13.6400 \tabularnewline
6  &  6.510  &  0.2398  &  0.1680  &   4.6300 \tabularnewline
7  &  6.566  &  0.3090  &  0.0227  &  12.6900 \tabularnewline
\hline
\end{tabular}\protect\caption{Sample 2: Elastic loss of initial coating with bond defects.}
\label{table:Sample2Loss} 
\end{center}
\end{table}

\begin{table}[t]
\begin{center}
\begin{tabular}{ccccc}\hline
Mode & Freq. & \multicolumn{3}{c}{Loss Angle $(\times 10^{-5}$)}\tabularnewline
 &  (kHz) & $\phi_{\mathrm{sample}}$ & $\phi_{\mathrm{substrate}}$& $\phi_{\mathrm{coating}}$ \tabularnewline
\hline 
1  &  1.077  &  0.0404  &  0.0069  &   1.5500 \tabularnewline
2  &  1.624  &  0.3734  &  0.0135  &  15.8000 \tabularnewline
3  &  2.473  &  0.0970  &  0.0391  &   3.0500 \tabularnewline
4  &  3.792  &  0.3122  &  0.0063  &  13.1100 \tabularnewline
6  &  6.536  &  0.1028  &  0.0664  &   2.3300 \tabularnewline
\hline
\end{tabular}\protect\caption{Sample 4: Elastic loss of initial coating with bond defects.}
\label{table:Sample4Loss} 
\end{center}
\end{table}

\begin{table}[!htbp]
\begin{center}
\begin{tabular}{ccccc}\hline
Mode & Freq. & \multicolumn{3}{c}{Loss Angle $(\times 10^{-5}$)}\tabularnewline
 & (kHz) & $\phi_{\mathrm{sample}}$ & $\phi_{\mathrm{substrate}}$& $\phi_{\mathrm{coating}}$ \tabularnewline
\hline 
1  &  1.071  &  0.0646  &  0.0069  &   2.3900 \tabularnewline
2  &  1.624  &  0.3689  &  0.0135  &  16.4700 \tabularnewline
3  &  2.472  &  0.0927  &  0.0391  &   3.2800 \tabularnewline
4  &  3.792  &  0.2947  &  0.0063  &  13.1800 \tabularnewline
6  &  6.530  &  0.1130  &  0.0664  &   3.5800 \tabularnewline
7  &  6.596  &  0.2554  &  0.0062  &  13.2200 \tabularnewline
\hline
\end{tabular}\protect\caption{Sample 4: Elastic loss of etched coating with bond defects removed.}
\label{table:Sample4EtchedLoss} 
\end{center}
\end{table}

\begin{figure}[!htbp]
    \centering
    \includegraphics[width=.45\textwidth]{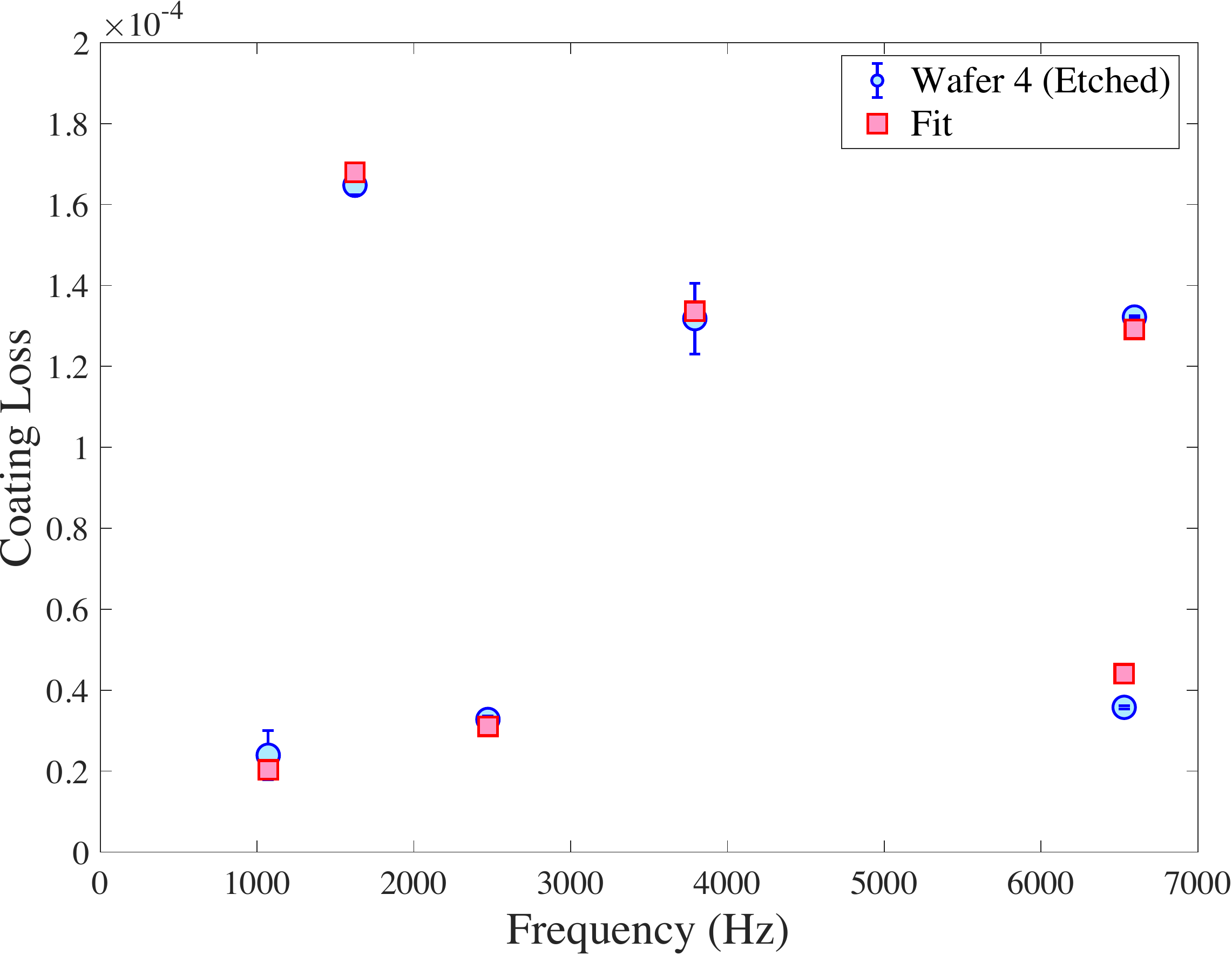}
    \caption{Coating loss $\phi_{\mathrm{coat}}$ for Sample 4 with the bond defects selectively etched away. A fit of bulk and shear losses yielded $\phi_{\mathrm{Bulk}} = (5.33 \pm 0.03)\times 10^{-4}$ and $\phi_{\mathrm{Shear}} = (0.0 \pm 5.2) \times 10^{-7}$.
}
     \label{fig:etchedLoss}
\end{figure}
The elastic loss of coated samples and the bare substrates are listed in Tables~\ref{table:Sample2Loss} \& \ref{table:Sample4Loss} for the initial samples (before etching). The elastic loss of the coating is calculated using Equation~\ref{eqn:phifromQ} and the $R$ values from Table~\ref{table:Rvalues}. Figure~\ref{fig:initialLoss} shows the coating loss for the initial samples as a function of modal frequency. The loss was sharply divided with the loss in the drumhead modes (2,4,7) being about $4\times$ higher than the loss in the radial modes (1,3,6). This pattern of loss bifurcation, which has been observed previously in amorphous coatings~\cite{abernathy} (except in that case the radial modes had higher loss), indicated a large difference in the loss from bulk and shear motion. From this observation, we chose to analyze the AlGaAs samples with a bulk and shear decomposition. 

We decomposed the coating loss into bulk and shear losses using the equation
\begin{equation}
\phi_{\mathrm{coating}} = R_{\mathrm{bulk}} \phi_{\mathrm{bulk}} + R_{\mathrm{shear}} \phi_{\mathrm{shear}},
\end{equation}
where the energy ratios, $R_{\mathrm{bulk}}$ and $R_{\mathrm{shear}}$, given in Table~\ref{table:Rvalues} were calculated using a finite element model programmed in COMSOL~\footnote{see www.comsol.com/comsol-multiphysics}.

\begin{figure}[!htbp]
    \centering
    \includegraphics[width=.45\textwidth]{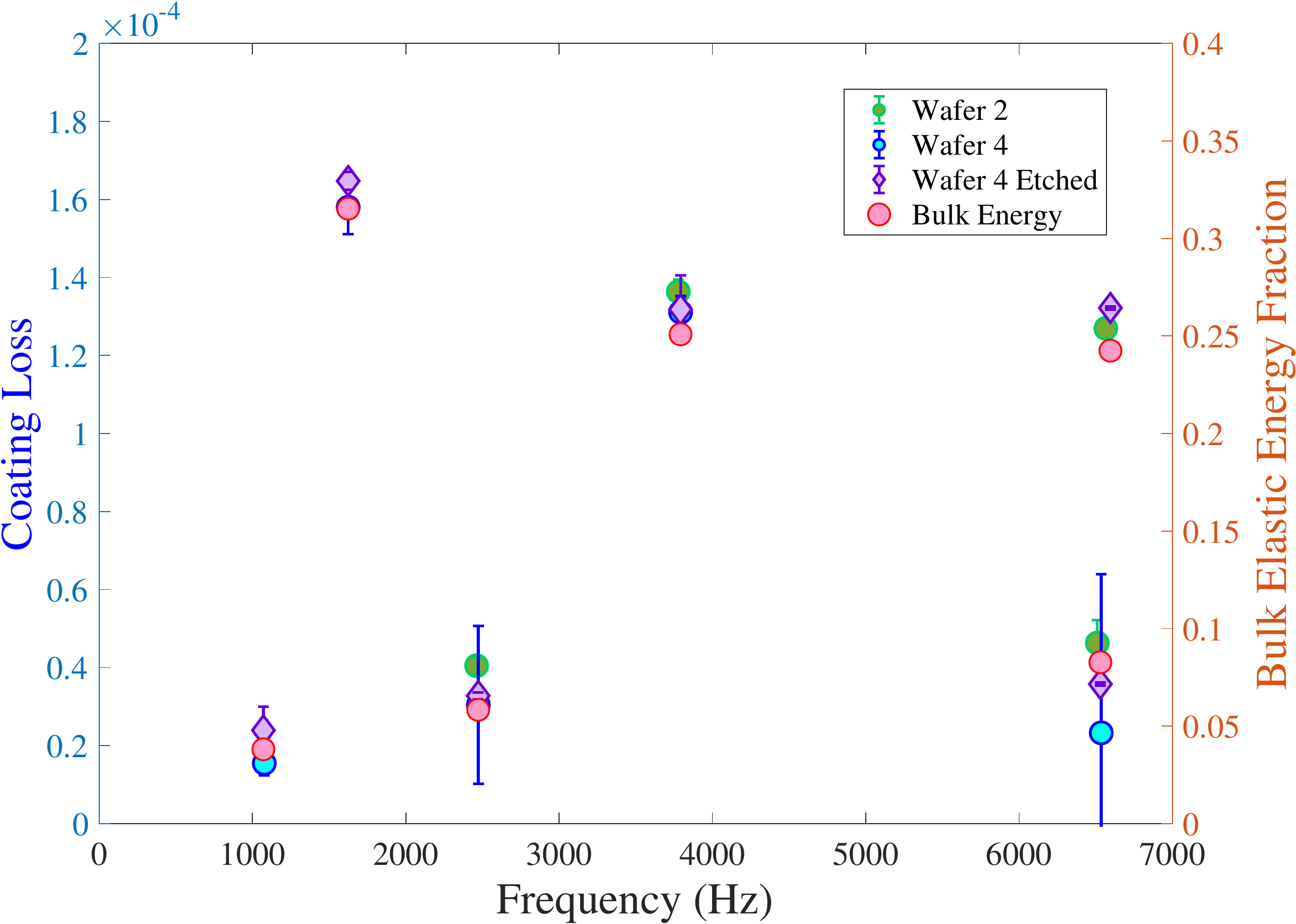}
    \caption{A comparison of variation of coating loss angle and the ratio of bulk stress energy to total stress energy versus modal frequency. The agreement illustrated the dominance of the bulk loss angle.}
    \label{fig:bulkLoss}
\end{figure}

As is shown in Figure~\ref{fig:bulkLoss}, the dependence of $R_{\mathrm{bulk}}$ with modal frequency was a good match with the dependence of $\phi_{\mathrm{coating}}$ versus frequency, indicating that the coating loss was dominated by the bulk loss. Indeed, when the data was fit there was no detectable contribution from $\phi_{\mathrm{shear}}$. The same analysis was performed on Sample 4 after the bond defects were removed by etching. The results are shown in Table~\ref{table:Sample4EtchedLoss} and Figure~\ref{fig:etchedLoss}. The initial coatings yield a bulk/shear loss of $\phi_{\mathrm{Bulk}} = (5.64 \pm 0.10)\times 10^{-4}$ and $\phi_{\mathrm{Shear}} = (0.0 \pm 1.5) \times 10^{-6}$, while the etched coating yielded a bulk/shear loss of $\phi_{\mathrm{Bulk}} = (5.33 \pm 0.03)\times 10^{-4}$ and $\phi_{\mathrm{Shear}} = (0.0 \pm 5.2) \times 10^{-7}$. If we attribute the difference in the pre-etch and the post-etch $\phi_{\mathrm{Bulk}}$ to the bond defects, then the bond defect loss only contributed about 5\% of the total coating loss.

To further test the validity of these results, we compared them with the coating loss calculated from the measured thermal noise of an optical cavity in Ref.~\cite{cole2013}. Here we built a finite element model of the 35-mm long cavity and, using a spot size of 0.25 mm (beam waist radius), calculated the bulk and shear energy ratios in the AlGaAs coatings to be $R_{\mathrm{bulk}} = 0.0898$ and $R_{\mathrm{shear}} = 0.9102$. Using these values we predict a cavity coating loss of $\phi_{\mathrm{coating}} = (4.78 \pm 0.05) \times 10^{-5}$, which agrees with the published result $\phi_{\mathrm{coating}} = (4 \pm 4) \times 10^{-5}$, albeit based on measurements from a sample with an 80-fold increased coating area (or ~20,000 times larger when compared with the optical spot size) than previously investigated.


\section{Conclusion}
Mechanical $Q$ measurements were performed both before and after defect removal on a large-area ($\approx$ 70-mm diameter) substrate-transferred AlGaAs--based crystalline coating. The coating elastic loss showed a 5\% reduction after defect removal, indicating that the loss contribution from the bond defects is, at most, small. A bulk/shear decomposition of the loss showed that the coating loss was due entirely to the bulk loss. The shear loss was on the order of $10^{-6}$ or less. This result suggests the intriguing possibility of minimizing the coating thermal noise by finding a configuration that maximizes the ratio of shear to bulk energy. With low thermal noise and excellent optical properties, GaAs/AlGaAs-based crystalline coatings are a promising candidate for future gravitational wave detectors and other precision optical applications.



Our research program will continue to investigate the source of the coating loss and methods to isolating the three loss angles for cubic crystals. On the large-area crystalline coating manufacturing front, optimization of the epitaxial growth and coating process, including the addition of post-growth polishing processes, are being investigated for the realization of defect-free crystalline coatings.

\section*{Acknowledgements}
We thank Martin Fejer, Markus Aspelmeyer, and the LIGO Scientific Collaboration Optics Working group for useful discussions and feedback. This paper has been assigned LIGO Document Number LIGO-P1800315. 

\section*{Funding} 
National Science Foundation (NSF) (PHY-0107417, 1707863, 1453252, 1611821).
\small{
\bibliography{Bibliography.bib}
\bibliographystyle{plain}
}

\end{document}